\documentclass[aip,jcp,reprint,groupedaddress]{revtex4-1}
\pdfoutput=1

\usepackage{amssymb}
\usepackage{graphicx}
 
\newcommand{\PBE}{{\mbox{\scriptsize PBE}}}

\newcommand{\G}{{\mbox{\scriptsize G}}}
\newcommand{\GO}{{\mbox{\scriptsize GO}}}
\newcommand{\Ox}{{\mbox{\scriptsize O}}}

\newcommand{\cx}{{\mbox{\scriptsize cx}}}

\newcommand{\qdp}{Quantum Device Physics Laboratory, Microtechnology and Nanoscience (MC2), Chalmers
University of Technology, SE-412 96 G{\"o}teborg, Sweden}
\begin{document}

\title{Graphene oxide and adsorption of chloroform: 
a density functional study}

\author{Elena Kuisma} %\affiliation{\qdp}
\author{C. Fredrik Hansson} %   \affiliation{\qdp}
\author{Th. Benjamin Lindberg}%  \affiliation{\qdp}
\author{Christoffer A. Gillberg}% \affiliation{\qdp}
\author{Sebastian Idh}      %\affiliation{\qdp}
\author{Elsebeth Schr{\"o}der} \email{schroder@chalmers.se}% \affiliation{\qdp}

\affiliation{\qdp}
\date{\today}
%%%%%%%%%%%%%%%%%%%%%%
\begin{abstract}
Chlorinated hydrocarbon compounds are of environmental concerns, 
since they are toxic to humans and other mammals, are widespread, 
and exposure is hard to avoid. 
Understanding and improving methods to reduce the amount of the
substances is important.
We present an atomic-scale calculational study of the adsorption of
chlorine-based substance chloroform (CHCl$_3$) on graphene oxide, as a step in 
 estimating the capacity of graphene oxide for filtering out such
substances, e.g., from drinking water. 
The calculations are based on density functional theory (DFT), and
the recently developed consistent-exchange functional for the
van der Waals density-functional method (vdW-DF-cx) is employed. 
We obtain values of the chloroform adsorption energy varying from roughly 
0.2 to 0.4~eV per molecule.
This is comparable to previously found results for chloroform adsorbed 
directly on clean graphene, using similar calculations.
In a wet environment, like filters for drinking water, 
the graphene will not stay clean and will likely oxidize, and thus
adsorption onto graphene oxide, rather than clean graphene, is a more 
relevant process to study.
\end{abstract}


\maketitle

%%%%%%%%%%%%%%%%%%%%%%
\section{Introduction}
\label{sec:intro}

Graphite oxide was first synthesized more than 150 years ago \cite{brodie}
but caught general 
interest\cite{stankovich06p282,stankovich06p155,schniepp06p8535,gomez07p3499,lerf,szabo06p2740,paci07,boukhvalov08p10697,mkhoyan09,dreyer09,Wan10,zhu10p3906,basu12p1,maliyekkal12p273,joshi14p752} 
only during the past few decades when 
research in 2D materials, in particular graphene, has started to bloom.
Graphite oxide is an alternative path to large-scale 
production of graphene, by liquid-phase exfoliation of graphite oxide 
into layers, called graphene oxide (GO), and subsequent reduction 
to graphene,\cite{stankovich06p282,stankovich06p155,schniepp06p8535} 
but already the GO sheets have intriguing and useful features.
GO can be understood as functionalized graphene, oxidized with
hydroxyl, epoxide and some carboxyl groups.
Its properties depend on the details of the oxidation: the type of, the number of
and the distribution of the functional groups.
GO has tuneable electric properties, obtained by changing the
functional groups, and with its thin size could be used for electronics.\cite{gomez07p3499}
GO is highly catalytic, highly solvable in water and 
other solvents, and is proposed for use as a gas sensor.\cite{basu12p1}

GO has been suggested as a material for use in
filtering of toxic compounds,\cite{maliyekkal12p273,joshi14p752} such as chlorinated 
hydrocarbon compounds. These are some of today's environmental concerns, 
since they are toxic to both humans and other mammals, and exposure is hard 
to avoid.  Exposure to chlorine-based compounds arises, 
e.g., from consumption of chlorinated drinking water or food supplies 
that have been contaminated by residues of industrial chemicals.\cite{Ros12,Ros06}  

We present a computational study of GO with state-of-the-art calculations,
using a recent implementation of density functional theory (DFT).
We study how GO  binds chloroform, one 
of the common chlorine-based substances, by calculating the binding energy
and its dependence on the structure of GO. 

GO has previously been studied\cite{lerf,szabo06p2740,paci07,boukhvalov08p10697,mkhoyan09,dreyer09}
in experiments and by use of calculational tools, including DFT. 
The GO itself is expected to be reasonably well described\cite{Wan10} by 
use of semilocal approximations of the exchange and correlation of DFT, 
such as in the PBE approximation,\cite{pebuer96}
but for our subsequent studies of chloroform physisorption 
it is imperative that the dispersive
nonlocal interactions  be included in a consistent way. 
Therefore we here use 
the van der Waals (vdW) density functional
method\cite{Dion,dionerratum,thonhauser,bearcoleluscthhy14,berland15p066501} (vdW-DF), 
in the vdW-DF-cx version,\cite{behy14,bearcoleluscthhy14}
for all calculations except for a comparison with
previous results, where we use PBE for some calculations.

Chloroform with graphitic or other carbon based materials was previously 
studied in a couple of experimental and computational 
studies.\cite{Ulb06,Fuj08,Ryb99,akesson12p174702,hooper08p891} 
Also, a DFT based study of ammonia adsorption was presented earlier.\cite{peng13}
However, to our knowledge there are no previously DFT studies of 
chloroform adsorption 
 on GO using methods that include the vdW interactions
consistently, such as here. 
Certainly, physisorption of chloroform on GO 
with the recent vdW-DF-cx has not previously been covered.  

This article is structured as follows: 
In Section~\ref{sec:methods} we describe the method of computation and the systems studied.
In Section~\ref{sec:results} the results are presented and discussed, along with a discussion
of the accuracy of our calculations, and 
Section~\ref{sec:conclusion} summarizes the study.   

%%%%%%%%%%%%%%%%%%%%%%
%%%%%%%%%%%%%%%%%%%%%%
\section{Physical system and computational method}
\label{sec:methods}
%%%%%%%%%%%%%%%%%%%%%%%
%%%%%%%%%%%%%%%%%%%%%%

In the following we describe GO and its functional groups
and how we model the relevant parts of GO in our adsorption
study. We also describe the methods used for the DFT
calculations and define the binding energies for the 
functional groups in GO and for chloroform on GO, 
including a discussion of what constitutes the zero 
point of the binding energies in these calculations.
This varies in the literature, making comparison difficult. 

\subsection{GO and chloroform}

GO has a graphene carbon network with an almost amorphous 
distribution of functional groups, and it is difficult to 
determine the types of functional groups present and their positions.
In Ref.~\onlinecite{lerf} Lerf et al.\ used reactions with
various reagents to supplement their previous NMR measurements.
They found that graphite oxide (and thus likely also GO) 
mainly has two types of functional groups: O bound in the
C-C bridge site, which is the 1,2-ether or epoxide,
and the C-OH or hydroxyl group.
They also found no support for O bound to two next-nearest
neighbor C atoms (the 1,3-ether) and very little for the 
carboxyl (the -COOH) group. 
Based on their measurements they put forward a structural model
for GO that has areas without functional groups (i.e., areas of 
clean graphene) and other areas with epoxide and hydroxyl groups
randomly distributed but close together. In their model
the carbon grid of GO is almost flat, except for the parts where
C atoms are attached to 
hydroxyl. The GO has functional groups on both sides of
the carbon grid and carboxyl groups are only present at the edges of GO.

In experiments,
fully oxidized GO is found to have a C:O ratio approximately 
2:1 or more.\cite{nakajima88p357}
However, GO is not always found in the fully oxidized state.
Even though the GO model of Lerf et al.\cite{lerf} has a relatively high concentration
of functional groups, the C:O ratio is only about 5:1 in the areas that are not
part of the GO sheet edges. 

In the present study of GO we consider structures with clusters of
functional groups on otherwise clean graphene, the clusters
being relatively small and disordered.
Thus, we compute the structures and energies involved in the
formation of epoxide and hydroxyl groups
on graphene for GO with low O concentration
(C:O ratio from 72:1 to 15:1). 
We use 
GO with functional groups on either both sides of the carbon grid (not symmetrically
positioned) or one side  with just a few clustered groups only. 
In our study we use periodic boundary conditions in space and thus the
GO has no edges. This means that according to the model by Lerf et al.\ there 
should not be any carboxyl groups included.

Chloroform (CHCl$_3$) consists of a central carbon atom with
three electronegative Cl atoms in a `tripod'
in one end and a H atom at the other end.
This gives chloroform a finite dipole moment, which affects its
physisorption properties.

In our chloroform adsorption study we initially studied 
mainly (but not exclusively) GO structures
that have all functional groups on the same side as the adsorbed molecule
because these groups were supposedly the ones that influence the 
chloroform-GO binding the most. However, we see that also groups on the
other side of the carbon grid affect the adsorption energy and our
study has been enlarged to enclose also systems with functional groups
on both sides.  

\subsection{Unit cell}
\label{sec:molecules}
%%%%%%%%%%%%%%%%%%%%%%
%%%%%%%%%%%%%%%%%%%%%%

\begin{figure}[bt]
\centering
\includegraphics[width=0.35\textwidth]{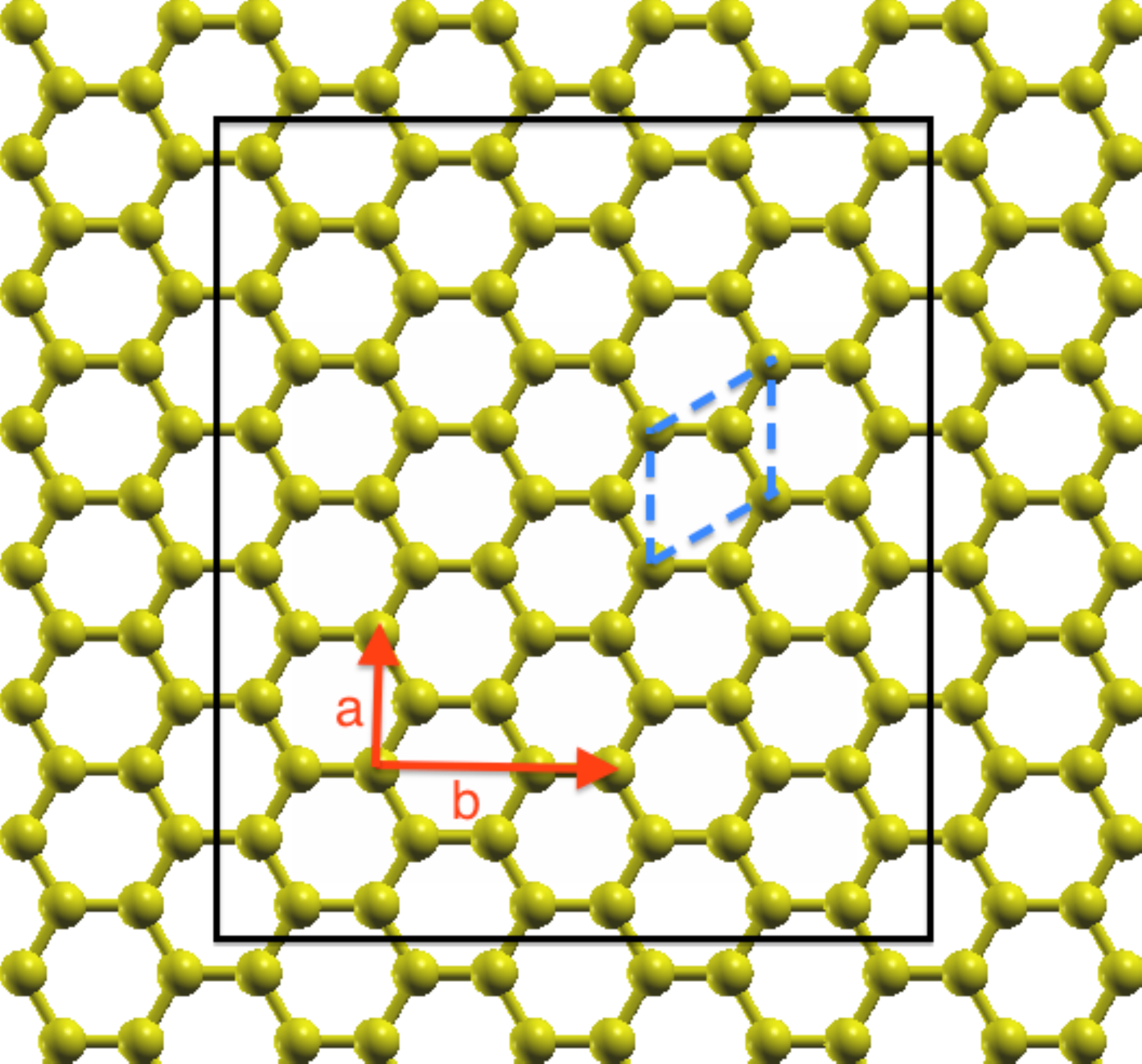}
\caption{Graphene slab. The unit cell 
(area delimited by the black solid line) used in this study consists 
of $3\times 6$ conventional rectangular graphene cells, each containing four 
carbon atoms, double the size of the primitive unit cell (blue dashed line). 
The lengths of unit vectors (red solid arrows) of the conventional cell 
are $a = 2.456$~{\AA} and $b = a\sqrt{3} = 4.254$~\AA. Visualization
(here and in Figures~\protect\ref{fgr:fgo-orient}, \protect\ref{fgr:fgoclvdwdfpc} 
and \protect\ref{fgr:fdip2}) using XCrySDen.\protect\cite{xcrysden}}
\label{fgr:slab}
\end{figure}

\begin{figure}[tb]
\centering
\includegraphics[width=0.30\textwidth]{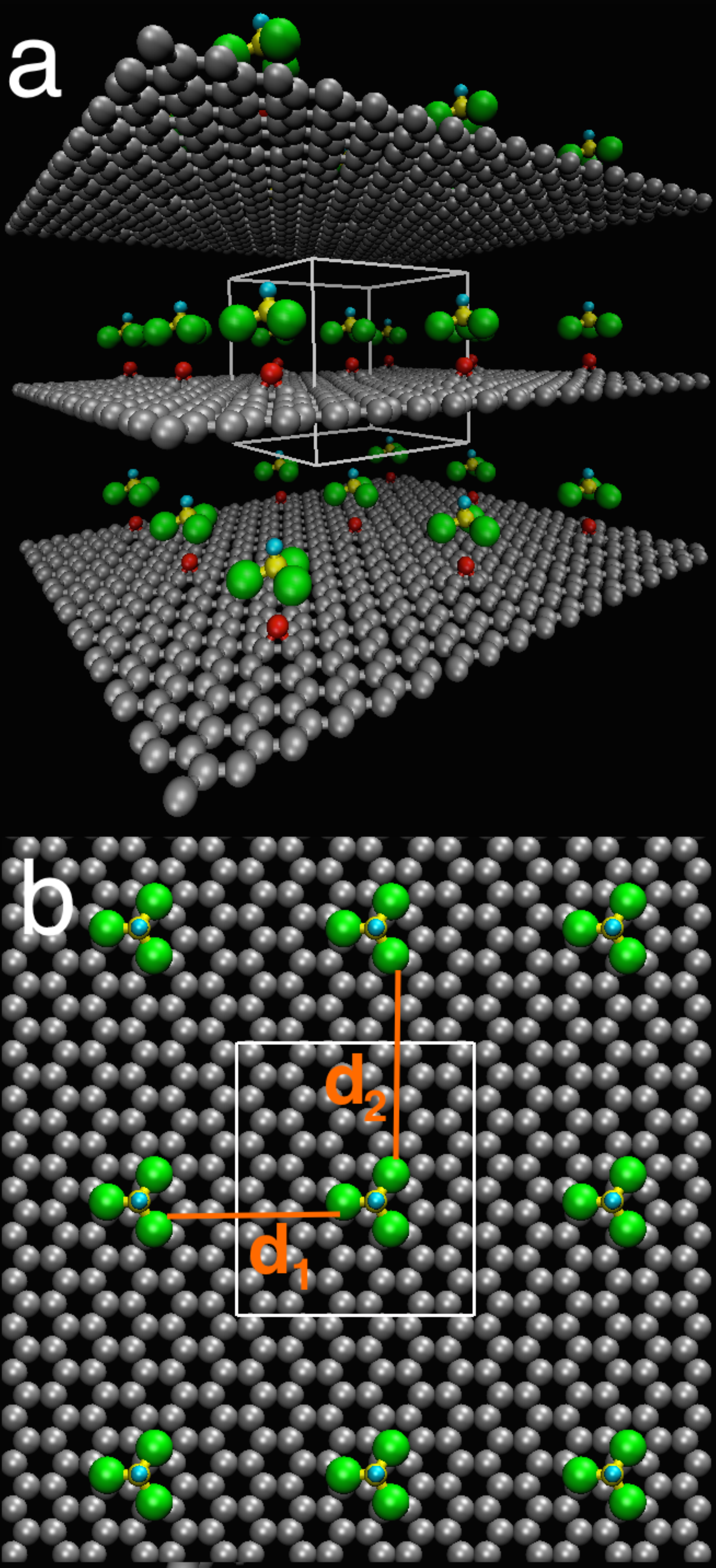}
\caption{Sketch of the unit cell used for calculations 
of chloroform adsorption on GO. 
The size of the unit cell (white box) is approximately 
$12.8\times14.8\times15.5$~\AA.  
\textit{Color legend:} Chloroform has yellow C; lime  Cl; blue H, and 
GO has gray C; red O. 
Closest atom-atom distances between chloroform molecules are indicated,
$d_1=10.4$~{\AA} %10.39 AA
and $d_2=11.9$~{\AA}. % 11.86 AA
Visualization using VMD.\protect\cite{vmd}}
 \label{fgr:supercell}
\end{figure}

To study GO 
we start out with a graphene slab with added functional epoxide and hydroxyl 
 groups. 
We use a $3\sqrt{3}\times 6$ orthorhombic unit cell with 72 graphene C atoms, 
as illustrated in Figure~\ref{fgr:slab}, with a C-C distance of 1.424~{\AA},
and periodic boundary conditions. 
The unit cell height is varied such that the amount of vacuum between each copy of the
system is  
approximately 10~{\AA}, thus for clean GO calculations the unit cell 
height is 10.5~{\AA}, and approximately 15.5~{\AA} for physisorption of chloroform on GO.
In the calculations of adsorbed chloroform the molecule-molecule nearest-neighbor distance is
12.8~{\AA} (the unit cell width) and the smallest lateral distance 
between any two atoms in neighboring chloroform molecules is 10.4~{\AA}, as 
illustrated in Figure~\ref{fgr:supercell}.

\subsection{Methods of computation}
\label{sec:calculations}
%%%%%%%%%%%%%%%%%%%%%%
%%%%%%%%%%%%%%%%%%%%%%

The formation of GO from graphene is a process in which hydroxyl and 
epoxide groups chemisorb on graphene. Such processes are expected to be 
well described 
with a semilocal density functional like PBE.\cite{pebuer96}
However, the adsorption of chloroform on GO is a physisorption process
and in such processes 
a robust description of the dispersive interactions is necessary.
We therefore perform the main DFT calculations using the vdW-DF 
method\cite{Dion,dionerratum,thonhauser,bearcoleluscthhy14,berland15p066501} 
in the consistent exchange
vdW-DF-cx version.\cite{behy14,bearcoleluscthhy14}
The vdW-DF-cx functional has been shown to work well for layered 
structures and aromatic molecules, and  
it accurately predicts the $a$ and $c$ lattice constants of solid 
graphite.\cite{bearcoleluscthhy14} 
It thus provides a balanced description between the chemical 
sp$^2$ bonding within the graphene sheets and 
the vdW interactions between layers and in physisorption. 

To describe the GO used as a substrate it is important that the method we use can also 
handle the balance between sp$^2$ and sp$^3$ binding, in the graphene 
patches and at the sites of the functional groups, respectively.
The fact that vdW-DF-cx shows reasonable results for the phase transistion between  diamond and
graphite, sp$^3$ and sp$^2$ materials, is encouraging.\cite{bearcoleluscthhy14}
To further document the ability of vdW-DF-cx for this problem we compare
formation energies for a number of unsaturated GO
configurations, obtained with both  the vdW-DF-cx and the semilocal PBE functional.

For the vdW-DF-cx results and most of the PBE results we
use {\sc Quantum Espresso}\cite{espresso,QEhttp} (QE). 
For further comparison we perform additional PBE calculations using the 
GPAW\cite{GPAW10,GPAWhttp} software.  
All calculations are carried out self-consistently.

In the QE calculations we use ultrasoft 
pseudo\-potentials\cite{vanderbilt90p7892,GBRV} 
with wavefunction and density cut-off energies 30 and 120~Ry, respectively. 
The force convergence threshold value is set to 2~meV{\AA}$^{-1}$, and 
the number of $\mathbf{k}$-points is 
$4\times 4 \times1$ (and $1\times 1 \times1$ for small molecules). 
With GPAW we use PAW setups\cite{PAW,mortensen05p035109} 
in the GPAW standard.\cite{gpawpaw}

The binding energy of the functional groups on graphene and of 
chloroform on GO, is calculated as
\begin{equation}
E_b= -(E_{AB} - E_{A} - E_{B}),
\label{eq:Eb}
\end{equation}
where $E_{AB}$ is the total energy of the full system and $E_{A}$ and $E_{B}$ 
are the total energies of its individual constituents 
(positive $E_b$ for systems that bind).
The zero point for the binding energy depends on the choice of 
individual constituents in subsystem A and B, e.g., whether to use
an isolated O atom or the O$_2$ molecule as one of the subsystems. 
These  choices  vary among authors of such studies.
Our choice is discussed below.

Spin polarization was not implemented in vdW-DF-cx at the time of our 
calculations, and thus we approximate
the binding energy of an epoxide on graphene as\cite{ziambaras07p155425,hamada14}
\begin{eqnarray}
E^\cx_b  &=      & -\left(E^\cx_\GO - E^\cx_\G - E^\cx_{^3\Ox}\right)\\
         &\approx& -\left(E^\cx_\GO - E^\cx_\G - \left(E^\PBE_{^3\Ox} 
         - E^\PBE_{^1\Ox} + E^\cx_{^1\Ox}\right)\right),
\label{eq:EbO}
\end{eqnarray}
where $E_{\GO}$ and $E_{\G}$ are the total energies of GO and graphene, 
and $E_{^3\Ox}$ and $E_{^1\Ox}$ are the energies of triplet ($^3$O) 
and singlet ($^1$O) O atoms, respectively, and
the superscript denotes the functional used, vdW-DF-cx or PBE. 
Thus, the difference in O singlet and triplet total energy 
in vdW-DF-cx is approximated by the same difference in PBE,
$E^{\cx}_{^3\Ox} - E^{\cx}_{^1\Ox}\approx E^{\PBE}_{^3\Ox} - E^{\PBE}_{^1\Ox}.$ 
The binding energies of the hydroxyl groups are calculated in a similar manner,
whereas the chloroform-on-GO calculations are carried out without considering spin,
with a non-spinpolarized calculation of chloroform and a graphene calculation
as the two individual constituents.

In the results section we report the height $h$ of the chloroform molecule above GO in
the adsorbed position. The height is taken as the  
projection in the $z$-direction 
(i.e., perpendicular to graphene) of the distance between 
the chloroform C atom and the nearest GO surface O atom. 
This means that an atom in chloroform may be closer to the GO O-atom than
the distance $h$.

%%%%%%%%%%%%%%%%%%%%%%
\section{Results and discussion}
\label{sec:results}
%%%%%%%%%%%%%%%%%%%%%%
%%%%%%%%%%%%%%%%%%%%%%
\subsection{Oxidized graphene}
\label{sec:go}
%%%%%%%%%%%%%%%%%%%%%%

\begin{figure*}[!htb]
 \centering
  \includegraphics[width=0.6\textwidth]{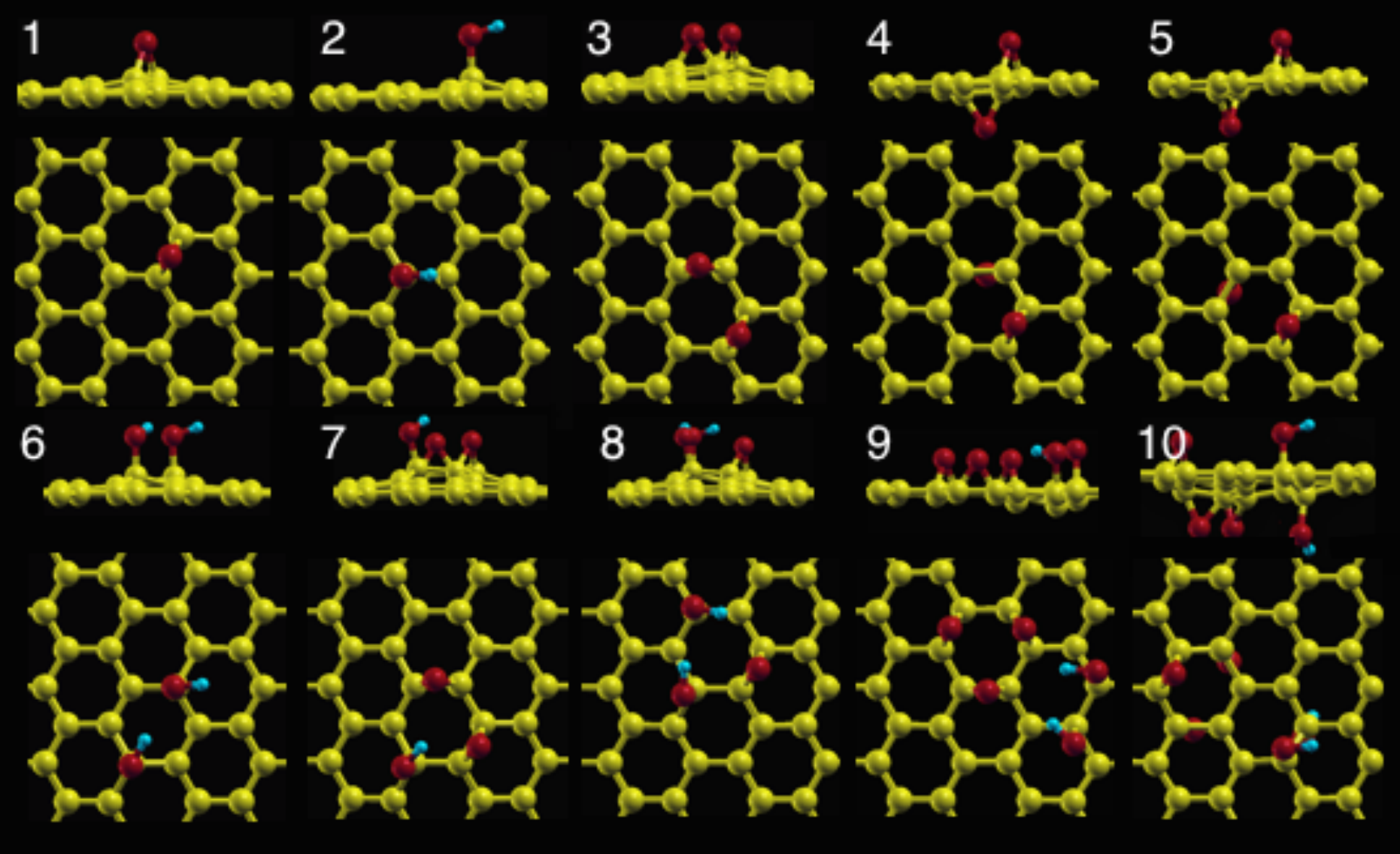}
  \hspace{1em}
  \includegraphics[width=0.24\textwidth]{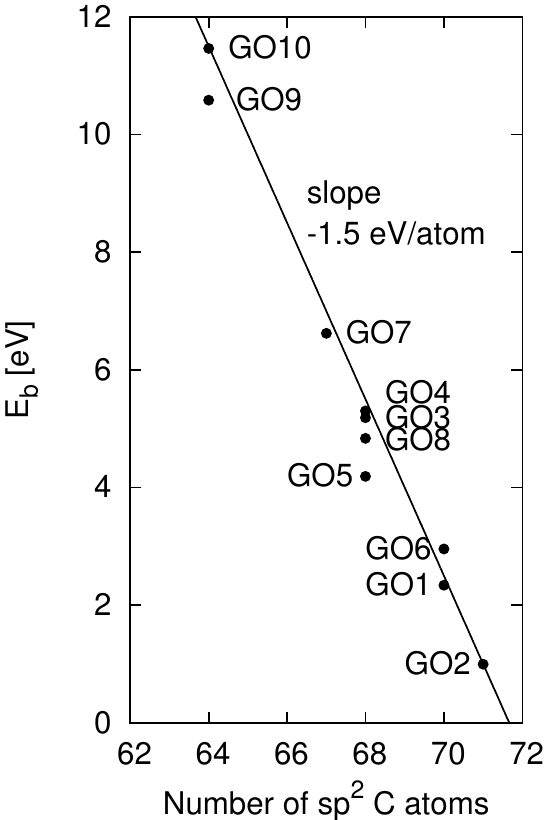}
  \caption{Left: Epoxide and hydroxyl groups on graphene. 
           The top and side views of optimized geometries of systems GO1--10. 
           Only a part of the unit cell is shown.
           \textit{Color legend for atoms:}
           yellow C; red O; blue H. 
           Right: The corresponding binding energies, as also 
           presented in Table~\protect\ref{tab:tgo-orient},
as a function of C atoms in the unit cell not bound to functional groups, i.e.,
number of remaining sp$^2$ C atoms of the initial 72 C atoms in the clean graphene. 
The method vdW-DF-cx is used for the structures shown here.  }
 \label{fgr:fgo-orient}
\end{figure*}

We focus on lightly oxidized graphene, with only a few functional 
groups per 72 C-atom unit cell.
We use
10 different configurations with up to three epoxide groups and up to two hydroxyl 
groups per unit cell, as illustrated in the left hand side of Figure~\ref{fgr:fgo-orient}. 
The functional groups are placed such that they form clusters, as expected
from experiments.\cite{lerf}
The resulting binding energies are presented in the graph on the right hand side
and in Table~\ref{tab:tgo-orient}. 
Structures GO1--3 and GO6--9 have groups on one side of the graphene
plane only, structures GO4--5 have one epoxide on each side, and structure
GO10 has epoxide and hydroxyl groups on both sides.  
The C--O lengths for the epoxides and hydroxyl groups are 
1.47~{\AA} and 1.52~{\AA} with the vdW-DF-cx functional. 

By studying the binding energy $E_b$ for each structure we find that
$E_b$ has an almost linear dependence on number of 
C atoms involved in binding the functional groups: the right hand side
of Figure~\ref{fgr:fgo-orient} shows that for every C atom involved 
(meaning there will be one less sp$^2$-bound C atom) 
$E_b$ increases by roughly 1.5 eV. For comparison, one single
hydroxyl group has a binding energy of about 1 eV and the epoxide group,
with its two C-O bonds, has a binding energy of about 2.3 eV, so the
gain in adding a functional group to a cluster is significantly higher, per C-O bond, than
just adding the group to a patch of clean graphene.
There is some spread in the
numbers, and the illustrations of the structures show that the 
clusters of functional groups are not all densely packed, leading to 
less gain in energy for sparse clusters than for more dense clusters.

The preference for having functional groups in clusters is seen already in a cluster of two
hydroxyl groups: we see a huge gain in energy (almost 1 eV) by pairing hydroxyl groups,
instead of having them separated. This is so even when the groups are on the same
side of the graphene grid and give rise to more distortion of the graphene
than the distortion created by one single hydroxyl group
(structure GO6 compared to GO2 times 2).  

More generally,  by visually
 comparing all of the structures in the left hand side of Figure~\ref{fgr:fgo-orient}
we find that structures with
groups on both sides (GO4, GO5, and GO10)
have a less distorted carbon structure than structures with groups
only on one side (all others), as expected.\cite{boukhvalov08p10697} 

Table \ref{tab:comparison} shows that for structures with two 
epoxides (GO3, GO4, and GO5) an energy gain can be obtained both
by clustering, with
gain per O atom 0.36 eV (from comparing GO1 and GO4),
and by having the epoxide groups on both sides of the graphene
instead of one side (gain 0.11 eV) 
but also that 
the largest gain is obtained by having the epoxides as nearest neighbors
as opposed to next-nearest neighbors (gain 1.21 eV). 
This is seen independent of DFT 
method (vdW-DF-cx and PBE) and computational code (QE and GPAW) and can also
be seen in literature values.\cite{slivancanin13p482}

Returning to the full set of formation energy results, Table~\ref{tab:tgo-orient},
we can compare the results of PBE calculations to vdW-DF-cx calculations.
The PBE calculations are expected to 
get both the sp$^2$ and sp$^3$ binding of the C atoms reasonably correct (but 
not so for the long-range interactions, which are important in chloroform physisorption).
We find that the vdW-DF-cx formation energies are systematically stronger than those
of PBE, with up to 23\% difference in formation energies.
However, the difference depends on the type of functional group(s) involved in
the GO structure: for epoxides the difference in formation energy of vdW-DF-cx
compared to PBE is 7--9\%, wheras the difference is 16--23\% for hydroxyl
groups, and mixed structures in the range 9--13\%.

It is important to note that these differences in vdW-DF-cx and PBE formation
energies \textit{include\/} changes to
the positions of the atoms when changing functional between vdW-DF-cx and PBE
and subsequently structually relaxing the atomic positions.
The hydroxyl functional group has an H atom pointing away from graphene,
this makes long-range interactions more relevant for hydroxyl that for
the epoxide. The single-bonded H atom also has a less stiff binding, and small 
changes in the forces on the atoms (from change of functional) 
can more easily move the H atoms than
the more stiffly bound O atoms. These differences between the hydroxyl and epoxide functional
groups are likely at least part of the reason for the larger PBE to vdW-DF-cx
energy difference when hydroxyl groups are involved: vdW-DF-cx may actually 
turn out to describe those groups better than PBE! However, without experiments
or high-quality quantum chemistry calculations to compare to we cannot claim 
that this is the case. 

\begin{table}[tb]
 \centering
  \caption{Binding energy $E_b$ and binding energy per O atom $E_b/$O 
of epoxide (O) and hydroxyl (OH) groups on graphene, 
calculated using the vdW-DF-cx and PBE functionals. 
Energies in units of eV.
See Figure~\ref{fgr:fgo-orient} for the systems GO\#. 
}
    \begin{ruledtabular}
  \begin{tabular}{llcccc} 
GO\# & GO struct.  &\(E^{\cx}_{b} \)&\(E^{\cx}_{b}/\mbox{O} \) &\(E^{\PBE}_{b} \) &\(E^{\PBE}_{b}/\mbox{O} \)  \\ 
\hline
GO1  &O & 2.341 &2.341& 2.166&2.166\\ 
GO2  &OH& 0.996 &0.996&0.769&0.769  \\ 
GO3  &2O& 5.187 &2.594&4.844&2.422\\ 
GO4  &2O& 5.301 &2.652 &4.937&2.469 \\ 
GO5  &2O& 4.190 &2.095&3.831 &1.916\\  
GO6  &2OH& 2.956 &1.478&2.495&1.247 \\ 
GO7  &2O, OH&6.620&2.207&6.042&2.014 \\ 
GO8  &O, 2OH&4.835&1.612&4.189&1.396 \\   
GO9  &3O, 2OH& 10.583&2.117&9.592&1.918 \\   
GO10 &3O, 2OH&11.462 &2.292&10.412&2.083 \\    
\end{tabular}
   \end{ruledtabular}
 \label{tab:tgo-orient}
\end{table}

\begin{table}[tb]
 \centering
  \caption{Binding energies $E_b$ and energy differences
for GO structures with two epoxide groups,
systems GO3, GO4 and GO5 and a single epoxide group, GO1. 
Calculations performed using QE and GPAW with the vdW-DF-cx and PBE functionals.
The structures differ by having epoxide on one (GO3) or both sides (GO4) of the graphene plane,
and by having the epoxides placed as nearest neighbors (GO4) 
or across the graphene carbon ring (GO5). 
The literature values (Ref.~\protect\onlinecite{slivancanin13p482}) are for similar
but not identical systems.
All energies in units of eV.}
     \begin{ruledtabular}
  \begin{tabular}{lcccc}
      &&This work&&\\
  \cline{2-4}   
  & QE &  QE & GPAW & Ref.\onlinecite{slivancanin13p482}\\
 Structure  & vdW-DF & PBE &  PBE &  PBE\\ 
 \hline
 GO3& 5.19 & 4.84 &  \\
 GO4& 5.30 & 4.94 & 5.15 & 4.76 \\ 
 GO5& 4.19 & 3.83 & 4.00 & 3.59 \\
\hline
GO4--GO3 & 0.11 & 0.10 & \\
GO4--GO5 & 1.21 & 1.11 & 1.15 & 1.17 \\
GO4-2$\times$GO1 & 0.72 & 0.61 & \\
 \end{tabular}
    \end{ruledtabular}
 \label{tab:comparison}
\end{table}

%%%%%%%%%%%%%%%%%%%%%%
\subsection{Adsorption of chloroform}
\label{sec:adsorption}

The main goal is to examine how chloroform binds to GO.
Our focus is on studying the effect on the binding energy 
of the presence of and the 
positions of the epoxide and hydroxyl 
groups and the relative orientation of the chloroform molecule.
Since GO is almost amorphous an exhaustive search is
prohibitive. Instead only the few functional groups closest to
the chloroform molecule are examined, keeping the rest of the unit 
cell clean of functional groups (i.e., using graphene), even if that might
result in a slightly worse binding energy than on fully oxidized GO. 
For the orientation of
chloroform we consider adsorption with the H atom pointing away
from the GO (``H up'') or towards the GO (``H down''), being aware that in the end 
positional relaxation due to the Hellmann-Feynman forces on the
atoms moves the groups and molecules to less well-defined orientations.
 
\begin{figure*}[!htbp]
 \centering
 \includegraphics[width=0.99\textwidth]{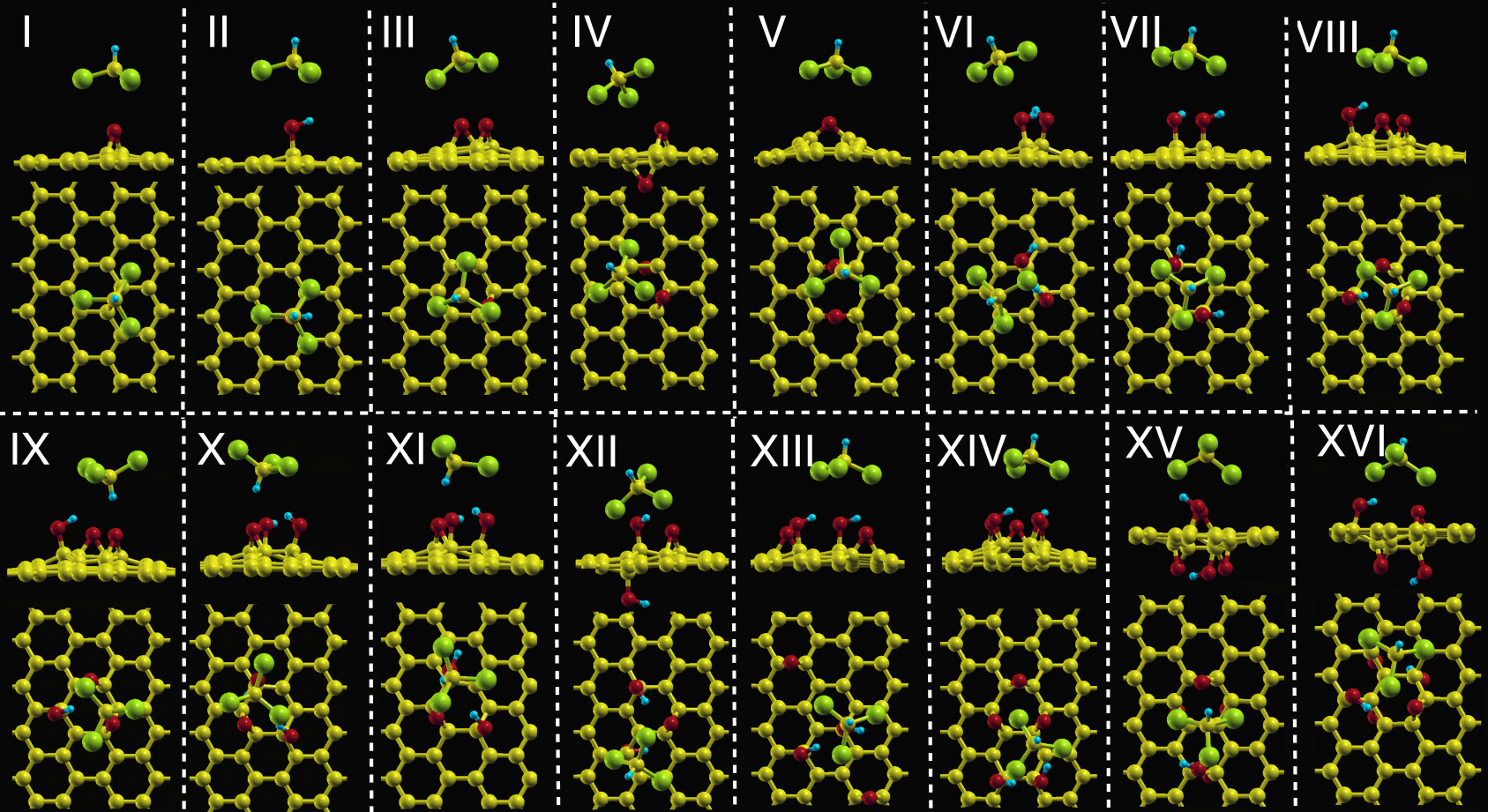} 
  \caption{Adsorption of chloroform on GO. 
Shown are top and side views of optimized geometries in systems I--XVI, obtained with
the method vdW-DF-cx..
Only a part of the unit cell is shown.
  The corresponding binding energies are presented in Table~\ref{tab:tgoclvdwdfpc}.
\textit{Color legend for atoms:}
 green Cl; yellow C; red O; blue H.  
}
 \label{fgr:fgoclvdwdfpc}
\end{figure*}

Several configurations of chloroform on GO are studied and data 
for 16 of these systems are presented in Figure~\ref{fgr:fgoclvdwdfpc}
and Table~\ref{tab:tgoclvdwdfpc}. 
The GO structures are mainly those presented in Figure \ref{fgr:fgo-orient}
and Table \ref{tab:tgo-orient} plus a few other.

Overall the adsorption energy lies approximately in the range 0.2--0.4 eV.
A closer look on the various adsorption systems compared for
similarities yields the following insights: I compared to II shows that adsorption
close to an (isolated) epoxide is more favorable by about 0.08 eV that 
adsorption close to (isolated) hydroxyl. The systems III, IV, and V  compared
to I show that adsorption close to a pair of epoxides 
is more favorable than adsorption on a single epoxide. The gain depends on
the relative positions of the epoxides, whether they are on the same side 
of the graphene plane (III and V) or not (IV), and whether the epoxides are
in nearest-neighbor positions (III) or sit across the C 6-ring (V). Placing the
epoxides on both sides of the graphene plane shows to be more favorable by
about 0.11 eV.

Systems VI and VII are used to examine the dependence on the relative positions of
hydroxyl pairs: in these calculations positioning the hydroxyl 
pair  across a C 6-ring is favored. Adsorption on two hydroxyl is also clearly 
preferable to adsorption on one hydroxyl only (II), by 0.20 eV.

For more complex systems of functional groups in the patch on graphene used
as a model of GO it is less clear which properties of the GO affect the adsorption
the most. We can, however, examine whether having the chloroform H atom pointing
towards or away from GO is favorable. For clean graphene it has previously been 
found that an orientation with H pointing away from graphene and the Cl-Cl-Cl tripod 
pointing towards graphene is 
most favorable, among the orientations considered, such as the Cl-Cl-H tripod 
pointing towards graphene, or the H atom
pointing towards graphene.\cite{akesson12p174702} 
Because of the more uneven structure of
the GO surface, compared to clean graphene with very little corrugation, 
we cannot distinguish the orientation  with the chloroform H atom
pointing to GO and the Cl atoms all pointing away to the orientation with the 
Cl-Cl-H tripod pointing to GO, we will therefore here only distinguish the
situation of chloroform H pointing mainly away from GO (``H up'')  
from H pointing mainly towards GO (``H down''). 

Systems VIII and IX differ in principle only by the orientation of chloroform
(besides the thus induced positional relaxations of both the GO and the chloroform
atoms). The adsorption energies of these systems indicate that ``H down'' 
is preferable. However, results for the more complex systems X, XI and XII 
clearly  show that there the ``H up'' orientation is more favorable, at least
if there are functional groups on both sides of the graphene plane, but also
that it matters how the Cl atoms are positioned relative to the atoms in epoxide.

Systems XIV, XV and XVI all have the same number of functional groups, 
placed either on one or both sides of the graphene plane. Again, the data show
that placing functional groups on both sides of the plane is preferable, even
when the number of functional groups are restricted and the number of 
functional groups close to chloroform thus becomes less, compared to having
all groups on the same side as the chloroform.

The measure $h$ in Table~\ref{tab:tgoclvdwdfpc} is an indication of the 
distance of chloroform from GO. Since GO is not flat, this measure is
may be both shorter or longer than for example the smallest atom-to-atom
distance in chloroform to GO. The $h$ measures the distance between 
the chloroform C atom and the closest GO O-atom, projected onto
the direction perpendicular to the underlying graphene net.
The values of $h$ in our calculations fall in the range 2.1 to 3.8 {\AA}.
These are reasonable distances for physisorption.

The adsorption of chloroform on clean graphene, without functional groups, 
was previously obtained\cite{akesson12p174702} as 0.36 eV 
using a similar method.
This value corresponds well to the most favorable configurations 
for chloroform on GO systems studied here. 

The information we obtain from studies like this of the physisorption of 
small molecules, like the binding 
energies and the orientation of chloroform in this study, may be useful
both as direct results, but also as input for modelling of larger systems.\cite{cooper12p34}

\begin{table}[tb]
 \centering
  \caption{Binding energies $E_b$ of chloroform on GO and height $h$
of chloroform above a GO O-atom 
(see Figure~\ref{fgr:fgoclvdwdfpc} for the GO-chloroform system numbers). 
GO\# refers to the GO structures of Figure~\ref{fgr:fgo-orient} and Table~\ref{tab:tgo-orient}. }
   \begin{ruledtabular}
  \begin{tabular}{llllrr}
System  & CCl$_3$H orient. &GO struct. &GO\# &$E_{b}$ [meV]& $h$ [\AA]\\ 
\hline
 I&H up&O&GO1&257 &3.34 \\ 
 II &H up&OH&GO2& 181 &3.45 \\   
 III &H up&2O &GO3& 225 & 3.56   \\ 
 IV &H up&2O&GO4& 382&2.72\\ 
 V &H up&2O &-&268 &3.40\\ 
 VI&H up&2OH& -&390&2.14 \\ 
 VII &H up& 2OH&GO6&199&3.69\\ 
 VIII &H up&2O, OH&GO7& 219 & 3.81 \\
 IX &H down&2O, OH&GO7& 286 &3.12  \\ 
 X &H down&O, 2OH&-& 319 & 2.99  \\ 
 XI &H down&O, 2OH&-& 296 & 2.99 \\ 
 XII &H up&O, 2OH&-& 422&2.44 \\ 
 XIII &H up& 2O, 2OH&-&243& 3.45 \\ 
 XIV &H up&3O, 2OH&GO9& 275&3.79 \\ 
 XV & H up&3O, 2OH&GO10& 333  &2.88\\ 
 XVI &H up&3O, 2OH&-& 391 &2.59  \\  
 \end{tabular}
  \end{ruledtabular}
 \label{tab:tgoclvdwdfpc}
\end{table}

\subsection{Accuracy}

\begin{figure}[bt]
\centering
\includegraphics[width=0.48\textwidth]{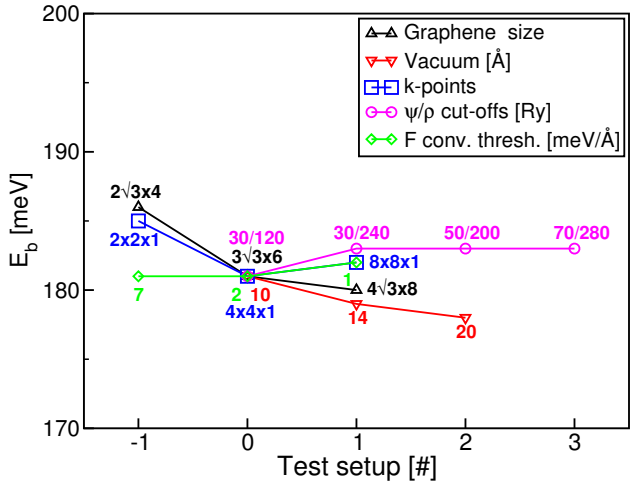}
\caption{Convergence tests of calculations
for chloroform on GO.
For these tests the chloroform-GO system
described as number II of
Figure~\ref{fgr:fgoclvdwdfpc} is used.
In this test all calculations use the same parameter values as the
production runs (``Test setup 0'') 
except for one parameter, the parameter indicated in the key of the figure.
Lines are for ease of identification.}
\label{fgr:ctest}
\end{figure}

We carefully check the convergence of parameters used in our    
QE calculations.
The parameter values are changed, one by one, to slightly better and
slightly worse values, and the binding energy for one of the
structures with chloroform on GO is calculated.
In Figure~\ref{fgr:ctest} we report the binding-energy dependence on
unit cell size, wavefunction and density cut-off energies,
the force convergence threshold value,
the number of $\mathbf{k}$-points, and the vacuum size.
The system studied is system II in Figure~\ref{fgr:fgoclvdwdfpc}.
The corresponding parameters are represented as a series of calculations,
in which the accuracy increases going from left to right in the figure.
The parameter values chosen from this convergence test for production runs
are more accurate than the default values of QE.
The convergence tests show that further improvements of the parameters
result in changes of adsorption energies of a few meV or less.

\begin{figure}[tb]
\centering
\includegraphics[width=0.35\textwidth]{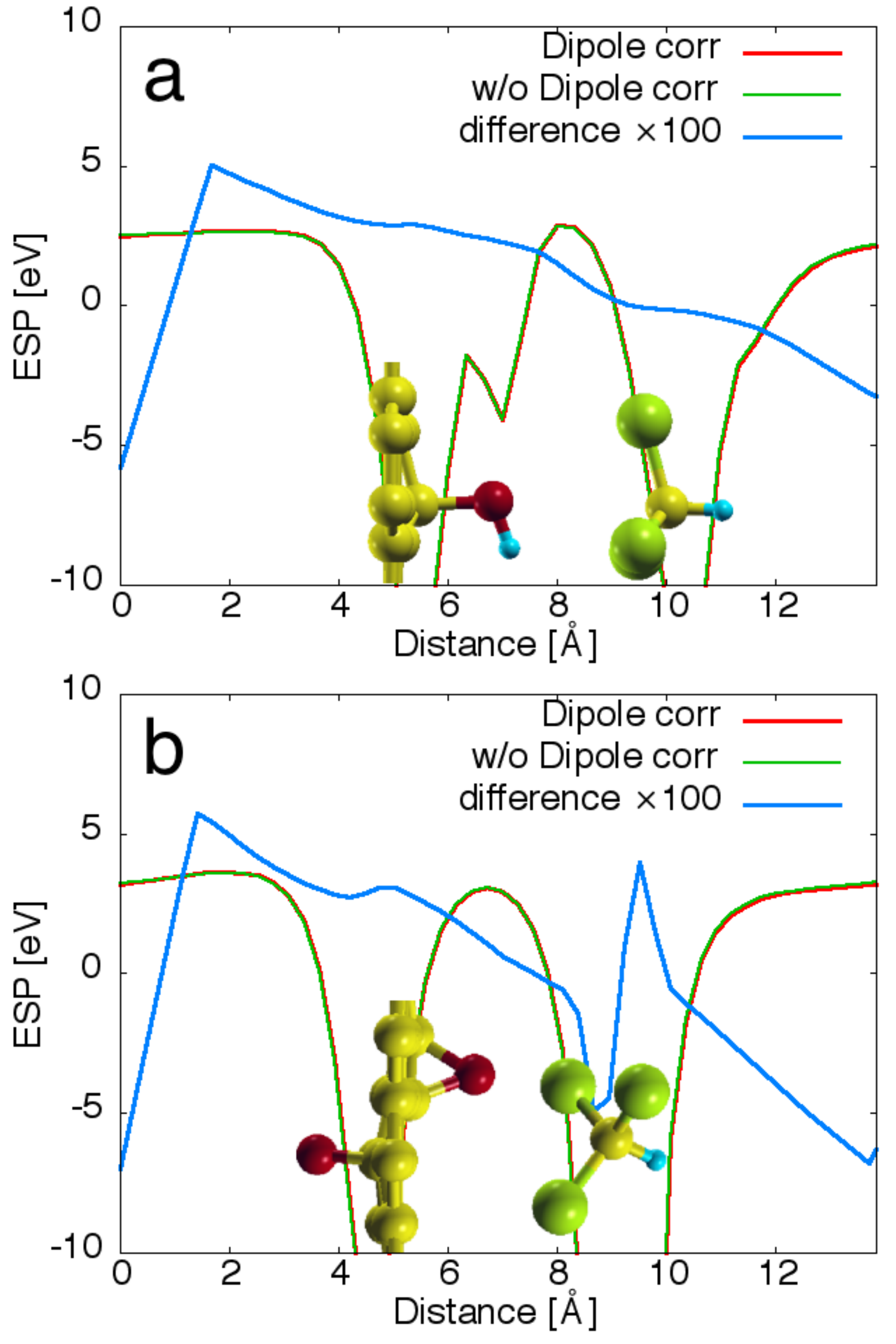}
\caption{Effect of the dipole correction
on the electrostatic potential in GO-chloroform systems.
The plot shows a potential line scan along the box $z$-axis through
the chloroform C atom (perpendicular to graphene) for the
systems  (a) II and (b) IV shown in Figure~\ref{fgr:fgoclvdwdfpc}.
To illustrate the length scales and positions of the constituents
a piece of the GO slab and the chloroform molecule are
placed at their position along the $z$-axes.
The unit cell lengths in these two calculations are 16~{\AA} (top)
and 14 {\AA} (bottom).
 Colors as in Figure \protect\ref{fgr:fgoclvdwdfpc}.
The solid blue line shows 100$\times$ the difference in potentials
with and without the dipole correction.}
\label{fgr:fdip2}
\end{figure}

Our calculations are carried out without correction for the dipole-dipole interaction
between two copies of the system in the $z$-direction.
The effect of a dipole correction on the systems is tested by
applying a dipole correction along the $z$-axis, for a couple of our
systems, in a manner described in Refs.\ \onlinecite{Ben99,Mey01}.
An example of a dipole correction test is presented in Figure~\ref{fgr:fdip2}.
The electrostatic potential (ESP) curves with and without dipole correction
are almost overlapping in the figures, and hence, for clarity  100$\times$ the
difference of these two curves is also plotted.
We see that the dipole correction only has minor effects on the (binding)
energy values of the studied systems.
This is not a trivial result, because the GO slab and chloroform are both polar objects.
However, the difference in binding energies with and without an applied
dipole correction is only 0.5--1.0~meV in most of the studied systems, with
one exception showing a less than $10$~meV difference.
    
%%%%%%%%%%%%%%%%%%%%%%
\section{Conclusions}
\label{sec:conclusion}

We present a computational study of chloroform physisorption on 
GO using DFT calculations with the vdW-DF-cx method.
The binding energy values vary from approximately 0.20 to 0.40~eV,
depending on the local environment on the GO, as well
as the orientation of the chloroform molecule relative to graphene
oxide.  
Thus we find that chloroform physisorbs rather strongly 
on GO 
and that graphene-oxide has potential as filtering 
material for chlorinated water. 

Further, we document the ability of vdW-DF-cx of balancing the sp$^2$ and sp$^3$
bindings of the GO C atoms. We do this by comparing the formation energy 
of the unsaturated GO structures, going from pure sp$^2$ binding in clean 
graphene to a mixture of sp$^2$ and sp$^3$ in the formation of GO from graphene.
The GO structures  are structurally relaxed with use of either 
the semilocal PBE or the vdW-DF-cx functionals. Besides a small offset in the formation energies
common to all the structures, we find the same formation energies for PBE and vdW-DF-cx calculations,
underlining the ability of vdW-DF-cx of handling the change from sp$^2$ to sp$^3$ bindings.
 
Our results for chloroform adsorption may be used as input data for modelling
larger systems, with or without thermodynamics, with or without additional molecules, such as water
or ions. While our study is not exhaustive 
in searching for all possible physisorption geometries we have included a number
of structures such that we have covered many of the relevant local environments
of GO functional groups.

%%%%%%%%%%%%%%%%%%%%%%
\begin{acknowledgments}
We thank Per Hyldgaard and Kristian Berland for providing access to 
a pre-release version of the vdW-DF-cx code with 
Quantum Espresso. 
The computations were performed on resources at Chalmers Centre for 
Computational Science and Engineering (C3SE) provided by the 
Swedish National Infrastructure for Computing (SNIC).
\end{acknowledgments}
%%%%%%%%%%%%%%%%%%%%%%
\bibliography{gocl-manus-v1.0ES}
%%%%%%%%%%%%%%%%%%%%%%%%%%%%%%%%%%%%%%%%%%%%%%%%%%%%%%%%%%%%%%%%%%%%%%%%
\end{document}